\documentclass[final,10pt,times,twocolumn]{elsarticle}
\usepackage[margin=1.8cm]{geometry}
\usepackage{amsmath}
\usepackage{amsfonts}
\usepackage{amssymb}
\usepackage{graphicx}
\usepackage{lmodern}
\usepackage{hyperref}
\usepackage{dcolumn}
\usepackage{bm}
\usepackage[mathlines]{lineno}
\usepackage{float}
\usepackage[T1]{fontenc}
\usepackage{subfigure}
\usepackage{mathtools, cuted}
\usepackage[english]{babel}
\usepackage{lipsum}

\journal{Physics Letters B}

\usepackage{numcompress}

\DeclareUnicodeCharacter{2212}{\ensuremath{-}}

\begin{document}
\newcolumntype{P}[1]{>{\centering\arraybackslash}p{#1}}
\title{\textbf{The other variants of mixed $\mu$-$\tau$ symmetry}}



\author[mymainaddress]{Pralay Chakraborty}
\ead{pralay@gauhati.ac.in}

\author[mymainaddress]{Subhankar Roy\corref{mycorrespondingauthor}}
\cortext[mycorrespondingauthor]{Corresponding author}
\ead{subhankar@gauhati.ac.in}

\address[mymainaddress]{Department of Physics, Gauhati University, India}

\begin{abstract}

Two new neutrino mass matrix textures exhibiting the \emph{mixed $\mu$-$\tau$ symmetry} are proposed. The mass matrices hint for a promising neutrino mixing schemes and find their connections with $\Delta(27)$ and $A_{4}$ discrete symmetry groups respectively. 
\end{abstract}
\maketitle

The standard model (SM) of particle physics\,\cite{Glashow:1961tr, Weinberg:1967tq, Herrero:1998eq} is well-tested in describing the masses and the interactions of the fundamental particles except for neutrino. The SM presents neutrino as a massless particle while the neutrino oscillation phenomenon\,\cite{Pontecorvo:1957cp} tells about the non-zero neutrino mass which is well tested in the experiments\,\cite{KamLAND:2002uet, SNO:2002tuh, Super-Kamiokande:1998uiq}. This indicates one has to go beyond the territory of SM to describe the tiny but non-zero mass of neutrino. In principle, the neutrino mass matrix $(M_\nu)$ originates from the Yukawa Lagrangian and it can be constructed from either Dimension-5 operator or see-saw mechanism. The neutrino mass matrix $M_\nu$ is an important tool to understand the hidden possibilities associated with the neutrino physics phenomenology. The $M_\nu$ being complex symmetric in nature, carries the information of twelve parameters: three mass eigenvalues $(m_1, m_2, m_3)$ three mixing angles ($\theta_{12}, \theta_{13}, \theta_{23}$), three CP phases $(\delta, \alpha, \beta)$ and three unphysical phases $(\phi_1, \phi_2, \phi_3)$. In most of the cases, the said unphysical phases are absorbed by redefining the three charged lepton fields. On the contrary, if these phases exist, they may contribute towards the texture of neutrino mass matrix. A detailed discussion is put forward in this regard in our earlier communication\,\cite{Dey:2022qpu}. A study on neutrino mass matrix texture centres round the reduction of the number of free parameters. In this regard, $\mu$-$\tau$ symmetry\,\cite{King:2018kka,Harrison:2002er}, texture zero\,\cite{Ludl:2014axa}, hybrid texture\cite{Singh:2018bap} and vanishing minor\,\cite{Lashin:2009yd,Araki:2012ip} are considered as certain specific ways to deal effectively with this problem. A particular pattern of neutrino mass matrix may carry the signature of certain underlying symmetry. The neutrino physics phenomenology includes different problems, and one such problem is whether neutrino is a Dirac or Majorana fermion is still unanswered\,\cite{Bilenky:2020vjk}. But the postulation of the Majorana nature of neutrinos offers several compelling theoretical advantages.\,\cite{Chakraborty:2022ess}. Here, in this work, we stick to the Majorana nature of neutrino which in turn leads to the symmetric nature of $M_{\nu}$. In order to gather the information of physical parameters, we need to diagonalize the $M_{\nu}$: $M^d = V^T.M_\nu.V$, where, $V$ is  known as Pontecorvo-Maki-Nakagawa-Sakata (PMNS) matrix. The $V$ carries all the parameters except the mass eigenvalues and this is testable in the experiments. The standard parametrization of $V$ is discussed in \ref{appendix 1}.

A new promising mixing scheme entitled \emph{mixed $\mu$-$\tau$ symmetry} is put forward in our earlier communication off late\,\cite{Dey:2022qpu}. The name suggests an inherent connection of the texture of the neutrino mass matrix with original $\mu$-$\tau$ symmetry. However, the former deals with CP phases, non-zero $\theta_{13}$, non-maximal $\theta_{23}$ and the arbitrary phases. A similar treatment in this line is seen in the mixing schemes such as $\mu$-$\tau$ reflection symmetry\,\cite{Xing:2022uax,Liu:2017frs}, $\mu$-$\tau$ anti-symmetry\,\cite{Xing:2015fdg,Xing:2020ijf} and phase deviated $\mu$-$\tau$ reflection symmetry\,\cite{Chamoun:2019pbh}. In earlier communication,\,\cite{Dey:2022qpu}, the title \emph{$\mu$-$\tau$ mixed symmetry} was attributed to a specific neutrino mass matrix texture. However, it is quite relevant to consider that such a name can correspond to a family of neutrino mass matrix textures, than a particular one. With this motivation, two new mixed $\mu$-$\tau$ symmetric textures as shown in the following are posited,

\begin{eqnarray}
\label{M1}
M^1_\nu &=&
\begin{bmatrix}
A &  B  &  - B^* \\
B &  D  &  F \\
-B^* & F &  D \\
\end{bmatrix},\\
\label{M2}
M^2_\nu &=&
\begin{bmatrix}
A &  B  &  -i\,B \\
B &  D  &  F \\
-i\,B & F &  D \\
\end{bmatrix},
\end{eqnarray}

where, $A,\,B,\,D$ and $F$ appearing in both the textures are complex parameters. Here, we wish to highlight that the neutrino mass matrix $M_{\nu}^{i}$ is considered as the general one in the sense, it shelters all twelve parameters including the unphysical phases. We demarcate $M_{\nu}^{i}$ from $\tilde{M}_{\nu}^{i}$ which excludes the unphysical phases. We see that,
\begin{equation}
M_\nu ^{i} = P^{*}. \tilde{M}_{\nu}^{i}. P^{*}.   
\end{equation}
where, $P= diag\,(e^{i\phi_1}, e^{i\phi_2}, e^{i\phi_3})$.

The texture of $M_{\nu}^1$ is associated with two constraint relations,

\begin{eqnarray}
(M^1_\nu)_{12}&=&-(M^1_\nu)_{13}^* \label{constraint 1 of M1},\\
(M^1_\nu)_{22}&=&(M^1_\nu)_{33}\label{constraint 2 of M1}.
\end{eqnarray}

The fact that the choice of the unphysical phases may influence the neutrino mass matrix texture leads to the following two relations concerning $M_{\nu}^{1}$,

\begin{eqnarray}
\phi_2 &=& \frac{1}{2}\left( Arg[(\tilde{M^1_{\nu}})_{12}]+Arg[(\tilde{M^1_{\nu}})_{13}]-2\phi_1+\pi\right)\nonumber\\&&+\frac{1}{4}\left( Arg[(\tilde{M^1_{\nu}})_{22}] - Arg[(\tilde{M^1_{\nu}})_{33}]\right),\\
\phi_3 &=& \frac{1}{2}\left(Arg[(\tilde{M^1_{\nu}})_{12}]+ Arg[(\tilde{M^1_{\nu}})_{13}]-2\phi_1+\pi\right)\nonumber\\&&-\frac{1}{4}\left(Arg[(\tilde{M^1_{\nu}})_{22}] - Arg[(\tilde{M^1_{\nu}})_{33}]\right).
\end{eqnarray}

\begin{table*}[!]
\centering
\begin{tabular}{P{2cm}P{3cm} P{3cm}}
\hline 
Parameters & Set-1 & Set-2\\
\hline
$m_1/eV$ & 0.0259 & 0.0259\\
\hline
$\theta_{12}/^\circ$ & 31.27-35.86 & 31.27-35.86\\
\hline
$\delta/^\circ$ & 268-272 & 269-271\\
\hline
$\phi_1/^\circ$ & 79.9-80.1 & - \\
\hline
$\phi_2/^\circ$ & 215.8-215.9 & 133.3\,-\,133.5\\
\hline
$\phi_3/^\circ$ & 30.7-30.9 &  130.4\,-\,130.6\\
\hline 
$\alpha/^\circ$ & 138.2-138.4 & 52\,-\,52.1\\
\hline
$\beta/^\circ$ & 105.7-105.9 & 69.2\,-\,69.4\\
\hline
\end{tabular}
\caption{The Set-1 and Set-2 corresponds to the numerical values of the inputs parameters for textures $M^1_\nu$ and $M^2_\nu$.} 
\label{Numerical Inputs}
\end{table*}

The constraints appearing in Eqs.\,(\ref{constraint 1 of M1})-(\ref{constraint 2 of M1}) bring forth four real independent equations of which two are shown as in the following,

\begin{eqnarray}
\frac{A_1}{\sin \theta_{13}}+B1&=&0 \label{first analytical expression of M1},\\
A_2 \tan \theta_{23}+B_2&=&0,\label{second analytical expression of M1}
\end{eqnarray}
where, $A_1$ and $B_1$ are expressed in terms of all physical and unphysical parameters excluding, $\theta_{13}$, whereas that $A_2$ and $B_2$ include all parameters except $\theta_{23}$\,(See \ref{appendix 1} for necessary details). 

Interestingly we see that the second constraint in Eq.\,(\ref{constraint 2 of M1})leads to two relations connecting the two mass ratios with other parameters,
 
\begin{eqnarray}
\label{massratio1}
\frac{m_1}{m_2}&=& \frac{A_3}{B_3},\\
\label{massratio2}
\frac{m_3}{m_2}&=& \frac{A_4}{B_4},
\end{eqnarray}

where, $A_{3}$, $B_{3}$, $A_{4}$ and $B_{4}$ are certain functions of the nine free parameters excluding the mass eigenvalues (For details, see \ref{appendix 1}).

We see that, for $M_{\nu}^2$, the following constraints are true,
\begin{eqnarray}
(M^2_\nu)_{12}&=& -i\,(M^2_\nu)_{13} \label{constraint 1 of M2},\\
(M^2_\nu)_{22}&=&(M^2_\nu)_{33}\label{constraint 2 of M2}.
\end{eqnarray}

As before, it is found that the above constraints hold good only when the following conditions are satisfied, 

\begin{eqnarray}
\phi_2-\phi_3 &=&\frac{\pi}{2}+ Arg[\tilde{(M^2_\nu)}_{12}]-Arg[\tilde{(M^2_\nu)}_{13}],\\
\phi_2-\phi_3 &=& \frac{1}{2}\left( Arg[\tilde{(M^2_\nu)}_{22}]-Arg[\tilde{(M^2_\nu)}_{33}\right).
\end{eqnarray}

From Eq.\,\ref{constraint 1 of M2}, we obtain the following two relations,

\begin{eqnarray}
\frac{A_5}{\sin \theta_{13}}+ B_5&=&0 \label{first analytical expression of M2},\\
A_6\tan \theta_{23}+ B_6&=&0 \label{second analytical expression of M2},
\end{eqnarray}

where, $A_{5}$ and $B_5$ are expressible in terms of eleven free parameters except $\theta_{13}$, whereas $A_{6}$ and $B_6$ carry the same parameters except $\theta_{23}$. The second constraint appearing in Eq.\,(\ref{constraint 2 of M2}) is common to both $M_{\nu}^{1}$ and $M_\nu^{2}$ and hence the expressions for mass ratios in terms of other parameters, appearing in Eqs.\,(\ref{massratio1}) and (\ref{massratio2}) are true for $M_{\nu}^2$ as well.

\begin{figure*}[!]
  \centering
    \subfigure[]{\includegraphics[width=0.24\textwidth]{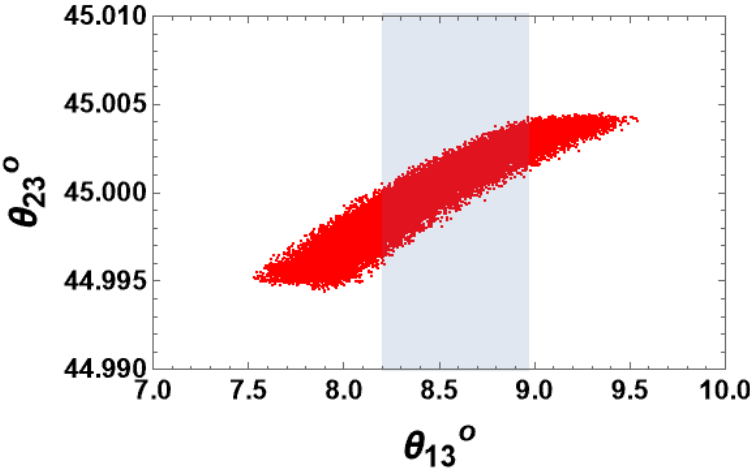}\label{fig:1(a)}} 
    \subfigure[]{\includegraphics[width=0.24\textwidth]{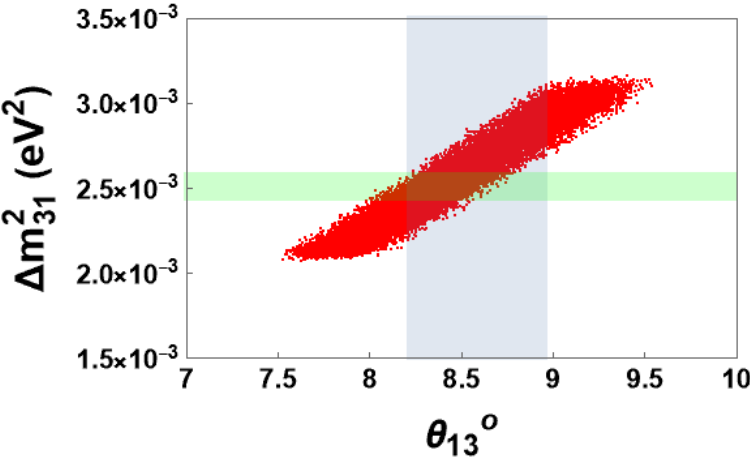}\label{fig:1(b)}} 
    \subfigure[]{\includegraphics[width=0.24\textwidth]{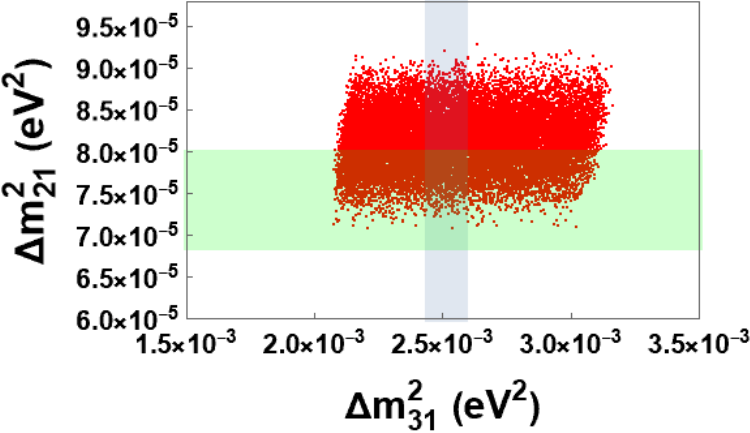}\label{fig:1(c)}}
    \subfigure[]{\includegraphics[width=0.24\textwidth]{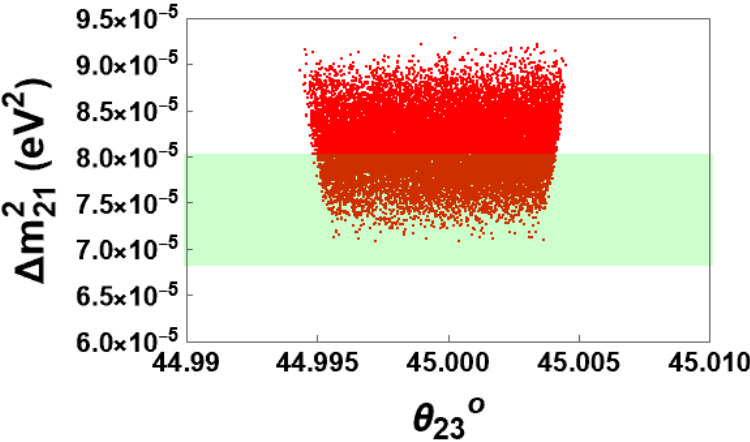}\label{fig:1(d)}}
    \caption{The correlation plots between (a) $\theta_{23}$ and $\theta_{13}$. (b) $\Delta m^2_{31}$ and $\theta_{13}$. (c)  $\Delta m^2_{31}$ and $\Delta m^2_{21}$ for normal ordering of neutrino masses. (d)  $\Delta m^2_{21}$ and $\theta_{23}$ for the texture $M^1_\nu$. The blue and the green strips represent the $3\sigma$ ranges of the concerned observational  parameters.}
\label{fig:1}
\end{figure*}

\begin{figure*}[!]
  \centering
    \subfigure[]{\includegraphics[width=0.24\textwidth]{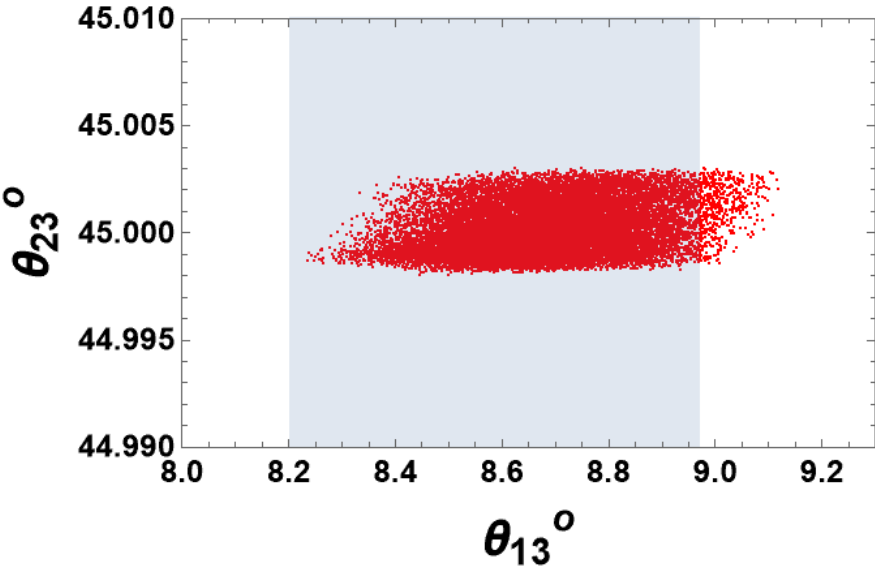}\label{fig:2(a)}} 
    \subfigure[]{\includegraphics[width=0.24\textwidth]{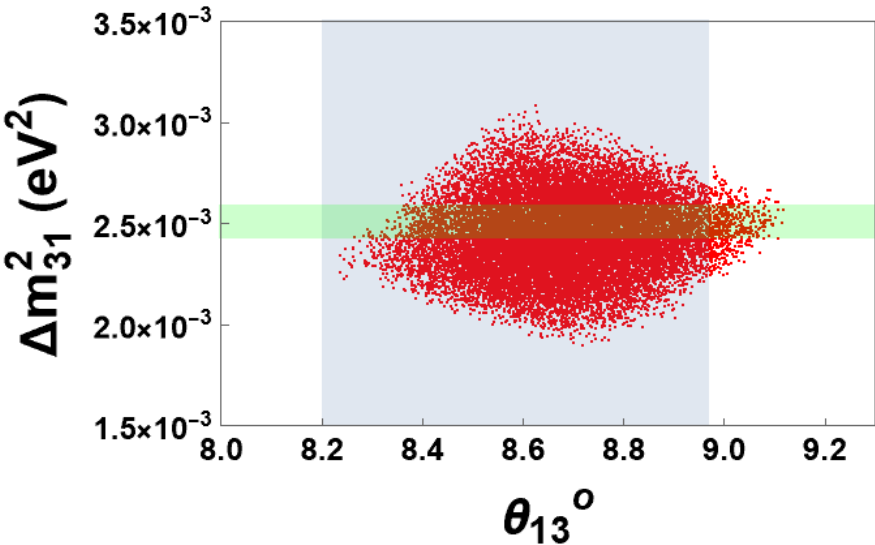}\label{fig:2(b)}} 
    \subfigure[]{\includegraphics[width=0.24\textwidth]{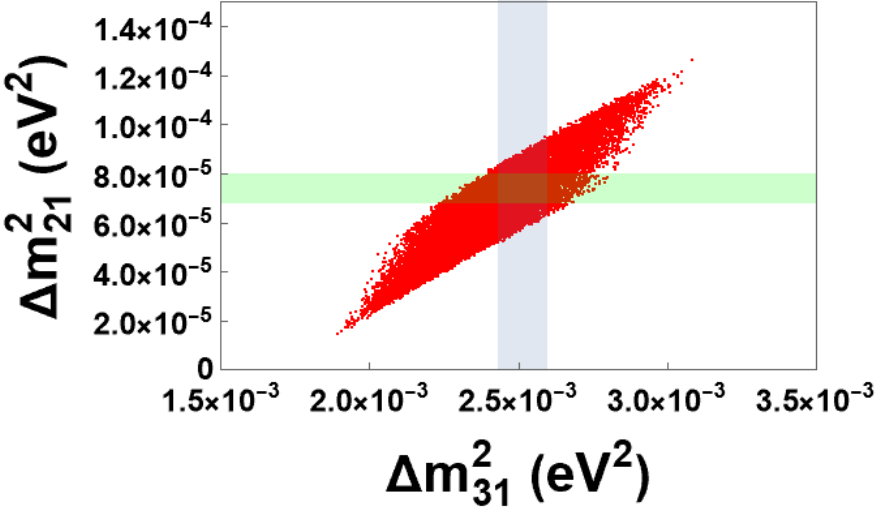}\label{fig:2(c)}}
    \subfigure[]{\includegraphics[width=0.24\textwidth]{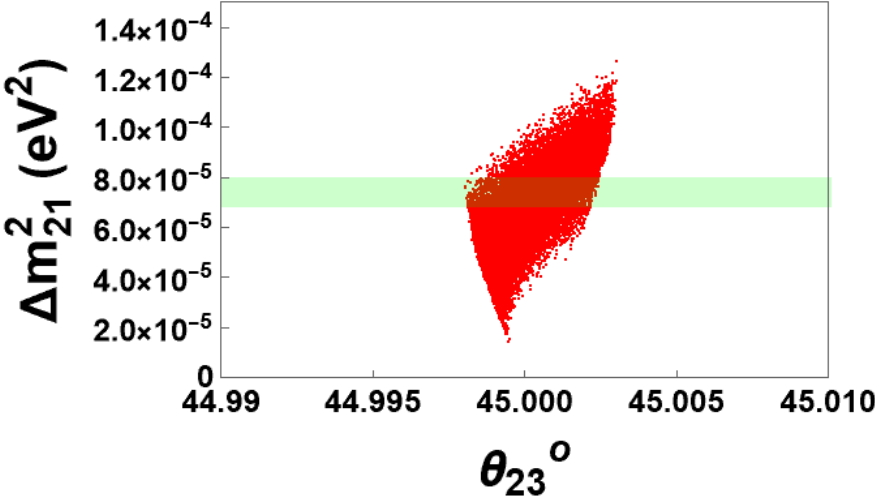}\label{fig:2(d)}}
    \caption{The correlation plots between (a) $\theta_{23}$ and $\theta_{13}$. (b) $\Delta m^2_{31}$ and $\theta_{13}$. (c)  $\Delta m^2_{31}$ and $\Delta m^2_{21}$ for normal ordering of neutrino masses. (d)  $\Delta m^2_{21}$ and $\theta_{23}$ for the texture $M^2_\nu$. The blue and the green strips represent the $3\sigma$ ranges of the concerned observational  parameters.}
\label{fig:2}
\end{figure*}

First, we discuss about the phenomenological implications of the texture $M_{\nu}^{1}$. We attempt to solve the four transcendental equations appearing in Eqs. (\ref{first analytical expression of M1})-(\ref{massratio2}). Sufficient random numbers for the observable parameters: $\Delta m^2_{21}$, $\Delta m^2_{31}$, $\theta_{13}$ and $\theta_{23}$ are generated and the correlation plots are obtained\,(see Figs \ref{fig:1(a)}-\ref{fig:1(d)}). From the plots, one deciphers that the mixing scheme is validated in the light of experimental observation \cite{Esteban:2020cvm, Gonzalez-Garcia:2021dve}. The input parameters needed for this analysis is shown in Table\,(\ref{Numerical Inputs}). Secondly, we do a similar analysis for $M_{\nu}^2$ and corresponding correlation plots are shown in Figs \ref{fig:2(a)}-\ref{fig:2(d)}. We see that the mixing pattern associated with $M_{\nu}^{2}$ is equally valid concerning the experimental results \cite{Esteban:2020cvm, Gonzalez-Garcia:2021dve}. For details of the input parameters, see Table\,(\ref{Numerical Inputs}). We mention that the present analysis is carried out in the light of normal ordering of neutrino masses with $m_1$ fixed at $0.0259$ eV.

Now, we shall try to see whether the proposed matrices in Eqs.\,(\ref{M1}) and (\ref{M2}) can be realised from the first principle or not. We first consider $M_{\nu}^1$. For this,  we extend the field content of the SM by introducing three right-handed neutrinos\,$(\nu_{e_R},\nu_{\mu_R},\nu_{\tau_R})$, a scalar singlet\,($\kappa$) and a scalar triplet\,($\Delta$). We associate the framework with $\Delta\,(27)$. The transformation properties of the field content associated with $\Delta\,(27)$ symmetry are summarised in Table\,\ref{Field Content of M1}. We construct the $SU(2)_L \times \Delta\,(27)$ invariant Lagrangian in the following way,

\begin{eqnarray}
- \mathcal{L}_Y &=& y_e (\bar{D}_{l_L}H)_{1_{00}}\,e_{R_{1_{00}}} + y_{\mu} (\bar{D}_{l_L}H)_{{1_{20}}} \,\mu_{R_{1_{10}}}+\nonumber\\&&\,y_{\tau}\,(\bar{D}_{l_L} H)_{{1_{10}}} \tau_{R_{1_{20}}} + y_1\,(\bar{D}_{l_L}\tilde{H})_{{1_{00}}}\,\nu_{e_{R_{1_{00}}}} \nonumber\\&&+ y_2 (\bar{D}_{l_L}\,\tilde{H})_{1_{10}} \nu_{\mu_{R_{1_{20}}}} + y_3 (\bar{D}_{l_L}\tilde{H})_{1_{20}}\,\nu_{\tau_{R_{1_{10}}}}\,\nonumber\\&&+\frac{1}{2}\,y_{R_1}\,(\bar{\nu}_{e_R}\,\nu^c_{e_R})_{1_{00}}\,\kappa_{1_{00}} + \frac{1}{2}\,y_{R_2}\,[(\bar{\nu}_{\mu_{R}}\,\nu^c_{\tau_{R}})\nonumber\\&&\,+ (\bar{\nu}_{\tau_{R}}\,\nu^c_{\mu_{R}})]_{1_{00}}\,{\kappa} +\, y_{T_2}\,\bar{D}_{l_{L_{3}}}\,( D_{l_L}^c\,{\Delta})_{3^*_{s_{1}}} \nonumber\\&&+\, h.c.
\label{Yukawa Lagrangian M1}
\end{eqnarray}

\begin{table*}[!]
\centering
\begin{tabular}{P{1.5cm}|P{0.5cm}P{2.5cm}P{2.5cm}P{0.3cm}P{0.3cm}P{0.3cm}} 
\hline
Fields & $D_{l_{L}}$ & $l_{R}$ & $\nu_{l_R}$ & $H$ & $\kappa$ & $\Delta$ \\ 
\hline
$SU(2)_{L}$ & 2 & 1 & 1 & 2 & 1 & 3 \\
\hline
$\Delta(27)$ & 3 & $(1_{00},1_{10},1_{20})$ & $(1_{00},1_{20},1_{10})$ & $3^*$ & $1_{00}$ & 3 \\
\hline
\end{tabular}
\caption{ The transformation properties of various fields under $SU(2)_L \times \Delta(27)$.} 
\label{Field Content of M1}
\end{table*}

The product rules under $\Delta\,(27)$ are given in \ref{appendix a}. We choose the complex vacuum alignment $\langle H \rangle_{\circ}=v_{H}\,(w,1,1)^{T}$ as per ref \,\cite{Branco:1983tn} and obtain the charged lepton mass matrix as shown below,

\begin{equation}
 M_{l} = v_{\phi}
 \begin{bmatrix}
 \omega^2 \, y_{e} & \omega^2 \,y_{\mu} & \omega^2 \, y_{\tau}\\
 y_{e} & \omega \,y_{\mu} & \omega^2 \,y_{\tau}\\
 y_{e} & \omega^2 \,y_{\mu} & \omega \,y_{\tau}\\
 \end{bmatrix},
 \label{chargrd lepton mass matrix}
\end{equation}
where, $\omega= e^{i\,2\pi/3}$, $\omega^2= \omega^*$.

The $M_l$ can be diagonalised as $M^{diag}_{l}=\,U^{\dagger}_{l_L}M_{l}U_{l_R}$, where, $M^{diag}_{l}=\,\sqrt{3}\,v_H\,diag\,(y_e,\,y_{\mu},\,y_{\tau})$. The $U_{l_L}$ and $U_{l_R}$ can  be expressed in the following way,
\begin{eqnarray}
 U_{l_L} &=& \frac{1}{\sqrt{3}}
 \begin{bmatrix}
 \omega^2\,e^{i\zeta} & e^{i\psi} & \omega\,e^{i\varphi}\\
 e^{i\zeta} & \omega^2\,e^{i\psi}  & \omega\,e^{i\varphi} \\
 e^{i\zeta} & e^{i\psi}  & e^{i\varphi} \\
 \end{bmatrix}, 
 \label{Mass matrix 6}\\
 U_{l_R} &=& diag( e^{i\zeta},\, \omega\,^{i\psi},\,\omega^2\, e^{i\varphi}).
\end{eqnarray}

It is to be noted that the inclusion of these arbitrary phases ($\zeta,\psi,\varphi$) in $U_{l_L}$ and $U_{l_R}$ has got many advantages and this idea is highlighted in ref\,\cite{Dey:2022qpu}. To experience the texture $M^1_\nu$, we set $\zeta=\frac{\pi}{2}$, $\psi=\frac{\pi}{6}$ and $\varphi=\frac{\pi}{2}$. 

On choosing the vacuum alignment $\langle H \rangle_{\circ}=v_{H}\,(w,1,1)^{T}$ and $\langle\kappa\rangle_{\circ}=v_{\kappa}$, the Dirac neutrino mass matrix ($M_D$) and right handed neutrino mass matrix ($M_R$) takes the following form,

\begin{eqnarray}
 M_{D} &=&
 \begin{bmatrix}
 y_1\,v_H\,\omega & y_2\,v_H\,\omega & y_3\,v_H\,\omega\\
 y_1\,v_H & \omega^2\,y_2\,v_H & \omega\,y_3\,v_H \\
 y_1\,v_H & \omega\,y_2\,v_H  & \omega^2\, y_3\,v_H \\
 \end{bmatrix},
 \label{Dirac Mass Matric M1}\\
 M_{R}&=&
 \begin{bmatrix}
 Y_{R_1}\,v_\kappa & 0 & 0\\
 0 & 0 & Y_{R_2}\,v_\kappa \\
 0 & Y_{R_2}\,v_\kappa  & 0 \\
 \end{bmatrix}.
 \label{M_R for M1}
\end{eqnarray}

The Type-I seesaw\,\cite{Branco:2020yvs, Brdar:2019iem} contribution towards the neutrino mass matrix appears as $M_{T_{1}}=\,-\,M_{D}M^{-1}_{R}M^{T}_{D}$. We take the choice of vacuum alignment $\langle\Delta\rangle_{\circ}=v_{\Delta}(1,1,1)^{T}$ for the $\Delta$ Higgs such that the Type-II \cite{SanchezVillamizar:2019fce} contribution takes the following form,

\begin{equation}
 M_{T_2} = t
 \begin{bmatrix}
 y_{T_2}\,v_\Delta &0 & 0\\
0 &  y_{T_2}\,v_\Delta & 0 \\
  0 &  0  & y_{T_2}\,v_\Delta \\
 \end{bmatrix}. 
 \label{Type-II for M1}
\end{equation}

 In the basis where the charged lepton mass matrix is non-diagonal\,(symmetry basis), the neutrino mass matrix is constructed as $M_{\nu_s}=M_{T_{1}}+ M_{T_{2}}$. Now, to shift to the basis where the charged lepton mass matrix is diagonal\,(flavour basis), we need to make a certain transformation ($U^{T}_{l_L} M_{\nu_s}  U_{l_L}$) and the effective neutrino mass matrix takes the following form,

\begin{equation}
M^1_\nu =
\begin{bmatrix}
A &  B  &  - B^* \\
B &  D  &  F \\
-B^* & F &  D \\
\end{bmatrix}.
\end{equation}

Now, we try to experience the texture $M^2_\nu$ from first principle. In this regard, we extend the field content of SM by adding three right handed neutrinos, a singlet scalar $\eta$, and a scalar triplet $\Delta$. In Table\,\ref{Field Content of M2}, we give the transformation properties of the field contents. The product rules under $A_4$ are given in \ref{appendix b}. The $SU(2)_L \times A_4$ invariant Lagrangian can be constructed as shown below,

\begin{eqnarray}
- \mathcal{L}_Y &=& y_e (\bar{D}_{l_L}H)_1\,e_{R_1} + y_{\mu} (\bar{D}_{l_L}H)_{1"} \,\mu_{R_{1'}}+\,y_{\tau}\nonumber\\&&\,(\bar{D}_{l_L} H)_{1'} \tau_{R_{1"}} + y_1\,(\bar{D}_{l_L}\tilde{H})_1\,\nu_{e_{R_1}} + y_2 (\bar{D}_{l_L}\nonumber\\&&\,\tilde{H})_{1'} \,\nu_{{\mu}_{R_{1''}}} + y_3(\bar{D}_{l_L}\tilde{H})_{1''}\,\nu_{{\tau}_{R_{1'}}}+\,\frac{1}{2}\,y_{R_1}\nonumber\\&&\,(\bar{\nu}_{e_R}\,\nu^c_{e_R})_1\,\eta_1 + \frac{1}{2}\,y_{R_2}\,[(\bar{\nu}_{\mu_{R}}\,\nu^c_{\tau_{R}})\,+ (\bar{\nu}_{\tau_{R}}\nonumber\\&&\,\nu^c_{\mu_{R}})]_1\,{\eta_1} +\, y_{T_2}\,(\bar{D}_{l_L}\, D_{l_L}^c)_{3_S}\,{\Delta}_3 +\, h.c.\nonumber\\
\label{Yukawa Lagrangian M2}
\end{eqnarray}

\begin{table*}[!]
\centering
\begin{tabular}{P{1.5cm}|P{0.5cm}P{2.5cm}P{2.5cm}P{0.3cm}P{0.3cm}P{0.3cm}} 
\hline
Fields & $D_{l_{L}}$ & $l_{R}$ & $\nu_{l_R}$ & $H$ & $\eta$ & $\Delta$ \\ 
\hline
$SU(2)_{L}$ & 2 & 1 & 1 & 2 & 1 & 3 \\
\hline
$A_4$ & 3 & $(1,1^{'},1^{"})$ & $(1,1^{"},1^{'})$ & $3$ & $1$ & 3 \\
\hline
\end{tabular}
\caption{ The transformation properties of various fields under $SU(2)_L \times A_4$.} 
\label{Field Content of M2}
\end{table*}

The charged lepton mass matrix under the choice of vacuum alignment $\langle H \rangle_{\circ}=v_{H}\,(1,1,1)^{T}$ \,\cite{Pramanick:2017wry} can be derived from Eq.\ref{Yukawa Lagrangian M2} as shown,

\begin{equation}
 M_{l} = v_{\phi}
 \begin{bmatrix}
  y_{e} &  y_{\mu} &  y_{\tau}\\
 y_{e} & \omega \,y_{\mu} & \omega^2 \,y_{\tau}\\
 y_{e} & \omega^2 \,y_{\mu} & \omega \,y_{\tau}\\
 \end{bmatrix},
 \label{chargrd lepton mass matrix}
\end{equation}

The left-handed and right-handed diagonalising matrix for $M_l$ are chosen in the following way,

\begin{eqnarray}
 U_{l_L} &=& \frac{1}{\sqrt{3}}
 \begin{bmatrix}
 e^{i\zeta} & \omega\, e^{i\psi} & \omega^2\,e^{i\varphi}\\
 e^{i\zeta} & \omega^2\,e^{i\psi}  & \omega\,e^{i\varphi} \\
 e^{i\zeta} & e^{i\psi}  & e^{i\varphi} \\
 \end{bmatrix}, 
 \label{Mass matrix 6}\\
 U_{l_R} &=& diag( e^{i\zeta},\, \omega\,^{i\psi},\,\omega^2\, e^{i\varphi}).
\end{eqnarray}

We choose the vacuum alignment $\langle H \rangle_{\circ}=v_{H}\,(1,1,1)^{T}$ and $\langle\eta\rangle_{\circ}=v_{\eta}$, to derive the Dirac neutrino mass matrix ($M_D$) and right handed neutrino mass matrix ($M_R$) as shown in the following,

\begin{eqnarray}
 M_{D} &=&
 \begin{bmatrix}
 y_1\,v_H\, & y_2\,v_H\, & y_3\,v_H\,\\
 y_1\,v_H & \omega^2\,y_2\,v_H & \omega\,y_3\,v_H \\
 y_1\,v_H & \omega\,y_2\,v_H  & \omega^2\,y_3\,v_H \\
 \end{bmatrix},
 \label{Dirac Mass Matric M1}\\
 M_{R}&=&
 \begin{bmatrix}
 Y_{R_1}\,v_\eta & 0 & 0\\
 0 & 0 & Y_{R_2}\,v_\eta \\
 0 & Y_{R_2}\,v_\eta  & 0 \\
 \end{bmatrix}.
 \label{M_R for M1}
\end{eqnarray}

The Type-II mass matrix can be derived from Eq.\ref{Yukawa Lagrangian M2} by considering the vacuum alignment $\langle\Delta\rangle_{\circ}=v_{\Delta}(0,1,-1)^{T}$\,\cite{Ma:2011yi} as follows,

\begin{equation}
 M_{T_2} = 
 \begin{bmatrix}
 0 &-y_{T_{2}}v_{\Delta} & y_{T_{2}}v_{\Delta}\\
-y_{T_{2}}v_{\Delta} &  0 & 0 \\
  y_{T_{2}}v_{\Delta} &  0  & 0 \\
 \end{bmatrix}. 
 \label{Mass matrix 5}
\end{equation}

In symmetry basis, the neutrino mass matrix $M_{\nu_s}$ is constructed by taking the contribution from Type-I and Type-II mechanisms. We have tried to derive the desired neutrino mass matrix in flavour basis as shown below,

\begin{equation}
M^2_\nu =
\begin{bmatrix}
A &  B  &  -i\,B \\
B &  D  &  F \\
-i\,B & F &  D \\
\end{bmatrix}
\end{equation}

To derive the texture $M^2_\nu$, we choose $\zeta=\frac{\pi}{6}$, $\psi=\frac{\pi}{2}$ and $\varphi=\frac{\pi}{3}$.

In summary, through this work, we have highlighted that the recently proposed \emph{mixed $\mu$-$\tau$ symmetry}\,\cite{Dey:2022qpu} can be extended as a family of neutrino mass matrix textures. In this line, the present work puts forward two new neutrino mass matrices which can be considered as two new members of this family. We have shown that both of these textures give promising lepton mixing schemes which fit well with the results from the neutrino oscillation experiments. We have shown that in the light of hybrid see-saw mechanism in association with $\Delta(27)$ or $A_4$, the respective  neutrino mass matrix textures can be obtained.

\section*{Acknowledgement}

The authors thank D Bhattacharjee, Department of Mathematics, Gauhati University for important suggestions. PC thanks M Dey, Gauhati University for fruitful discussions. The research work of PC is supported by the Innovation in Science Pursuit for Inspired Research (INSPIRE), Department of Science and Technology, Government of India, New Delhi vide grant No. IF190651. 
 
\biboptions{sort&compress}
\bibliography{ref.bib}

\appendix

\section{Parametrization of PMNS Matrix \label{appendix 1}}

The matrix $V$ is parametrised using nine parameters as $V=P_1.U.P_2$, where $P_1=diag(e^{i \phi_1},e^{i \phi_2},e^{i \phi_3})$ carries the unphysical phases and $P_2=diag(e^{i \alpha},e^{i \beta}, 1)$ carries the two Majorana phases. The Particle Data Group (PDG)\cite{ParticleDataGroup:2020ssz} has adopted   a parametrisation for $U$ termed as standard parametrisation as shown below,

\begin{eqnarray}
U &=& \begin{bmatrix}
1 & 0 & 0\\
0 & c_{23} & s_{23}\\
0 & - s_{23} & c_{23}
\end{bmatrix}\times \begin{bmatrix}
c_{13} & 0 & s_{13}\,e^{-i\delta}\\
0 & 1 & 0\\
-s_{13} e^{i\delta} & 0 & c_{13}
\end{bmatrix}\nonumber\\
&& \quad\quad\times\begin{bmatrix}
c_{12} & s_{12} & 0\\
-s_{12} & c_{12} & 0\\
0 & 0 & 1
\end{bmatrix},
\end{eqnarray}

where, $c_{ij}=\cos \theta_{ij}$ and $s_{ij}=\sin \theta_{ij}$. The matrix $U$ is observable in the oscillation experiments. However the latter keeps silent on the prediction of two Majorana phases.

\section{Product Rules of $\Delta(27)$ \label{appendix a}}

The non-Abelian discrete group $\Delta\,(27)$ is a subgroup of $SU(3)$ has 27 elements which can be divided into 11 equivalence classes \cite{Luhn:2007uq}. The group has two three dimensional representations and nine one-dimensional representations.

The group multiplication rules are shown below,

\begin{eqnarray}
3 \times 3&=&3^*_{S_1}+3^*_{S_2}+3^*_{A},\\
3 \times 3^* &=& \sum_{r=0}^{2} 1_{r,0}+\sum_{r=0}^{2} 1_{r,1} \nonumber\\&& +\sum_{r=0}^{2} 1_{r,2},\\
1_{r,p}\times 1_{r', p'}&=&1_{(r+r') \,\text{mod}\,3,\,\, (p+p') \,\text{mod}\, 3},\\
\end{eqnarray}

where,

\begin{eqnarray}
(3\times3)_{3^*_{S_1}}&=&\begin{bmatrix}
a_1 b_1 \\
a_2 b_2\\
a_3 b_3\\
\end{bmatrix},\\
(3\times3)_{3^*_{S_2}}&=&\frac{1}{2}\begin{bmatrix}
a_2 b_3+a_3 b_2\\
a_3 b_1+a_1 b_3\\
a_1 b_2+a_2 b_1\\
\end{bmatrix},\\
(3\times3)_{3^*_{A}}&=&\frac{1}{2}\begin{bmatrix}
a_2 b_3-a_3 b_2\\
a_3 b_1-a_1 b_3\\
a_1 b_2-a_2 b_1\\
\end{bmatrix},
\end{eqnarray}

and,

\begin{eqnarray}
1_{00}&=&(a_1\bar{b_1}+a_2\bar{b_2}+a_3\bar{b_3})\\
1_{10}&=&(a_1\bar{b_1}+\omega^2 a_2\bar{b_2}+\omega a_3\bar{b_3})\\
1_{20}&=&(a_1\bar{b_1}+\omega a_2\bar{b_2}+\omega^2 a_3\bar{b_3})\\
1_{01}&=&(a_1\bar{b_2}+ a_2\bar{b_3}+a_3\bar{b_1})\\
1_{11}&=&(a_1\bar{b_2}+\omega^2 a_2\bar{b_3}+\omega a_3\bar{b_1})\\
1_{21}&=&(a_1\bar{b_2}+\omega a_2\bar{b_3}+\omega^2 a_3\bar{b_1})\\
1_{02}&=&(a_1\bar{b_3}+ a_2\bar{b_1}+a_3\bar{b_2})\\
1_{12}&=&(a_1\bar{b_3}+ \omega^2 a_2\bar{b_1}+\omega a_3\bar{b_2})\\
1_{22}&=&(a_1\bar{b_3}+ \omega a_2\bar{b_1}+\omega^2 a_3\bar{b_2})
\end{eqnarray}

\section{Product Rules of $A_4$ \label{appendix b}}

The discrete group $A_4$ is a subgroup of $SU(3)$ and it has has 12 elements. The group has four irreducible representation: three one dimensional representations and one of dimension three. In S-basis \cite{Altarelli:2010gt}, the product of two triplets gives,

\begin{equation}
3\times3=1+1'+1"+3_S+3_A,
\end{equation}

where,

\begin{eqnarray}
1&=&(a_1\bar{b_1}+a_2\bar{b_2}+a_3\bar{b_3})\\
1'&=&(a_1\bar{b_1}+\omega^2 a_2\bar{b_2}+\omega a_3\bar{b_3})\\
1"&=&(a_1\bar{b_1}+\omega a_2\bar{b_2}+\omega^2 a_3\bar{b_3})\\
(3\times3)_S&=&\begin{bmatrix}
a_2 b_3+a_3 b_2\\
a_3 b_1+a_1 b_3\\
a_1 b_2+a_2 b_1\\
\end{bmatrix},\\
(3\times3)_A&=&\begin{bmatrix}
a_2 b_3-a_3 b_2\\
a_3 b_1-a_1 b_3\\
a_1 b_2-a_2 b_1\\
\end{bmatrix}.
\end{eqnarray}

The trivial singlet can be obtained from the following singlet product rules,

\begin{equation}
1\times1=1,\,\,1'\times1"=1,\,\,1"\times1'=1.
\end{equation}

\section{Analytical Expressions \label{appendix 2}}

The expressions for $A_1$, $B_1$, $A_2$, $B_2$, $A_3$, $B_3$, $A_4$, $B_4$, $A_5$, $B_5$, $A_6$, and $B_6$ appearing in Eqs. (\ref{first analytical expression of M1})-(\ref{second analytical expression of M2}) are shown below,

\begin{eqnarray}
A_1 &=& \frac{1}{2} \sin 2\theta_{12}\big[b \cos \theta_{23} \sin (2\alpha+\phi_1+\phi_2)-\cos \theta_{23}\nonumber\\&& \sin (2\beta+\phi_1+\phi_2)+b\sin \theta_{23} \sin (2\alpha+\phi_1+\phi_3) \nonumber\\&&-\sin \theta_{23}\sin (2\beta+\phi_1+\phi_3)\big],
\end{eqnarray}
\begin{eqnarray}
B_1 &=& a \sin \theta_{23} \sin(\delta-\phi_1-\phi_2)+b \cos^2 \theta_{12}\,\sin \theta_{23} \nonumber\\&&\sin(2\alpha+\delta+\phi_1+\phi_2)+\sin^2 \theta_{12} \sin \theta_{23} \sin(2\beta\nonumber\\&&+\delta+\phi_1+\phi_2)-a \cos \theta_{23} \sin(\delta-\phi_1-\phi_3)-\nonumber\\&&b \cos^2 \theta_{12}\cos \theta_{23}\sin(2\alpha+\delta+\phi_1+\phi_3)-\cos \theta_{23}\nonumber\\&&\sin^2 \theta_{12}\sin(2\beta+\delta+\phi_1+\phi_3),
\end{eqnarray}
\begin{eqnarray}
A_2 &=& a \sin \theta_{13} \cos(\delta-\phi_1-\phi_2)-b \cos^2 \theta_{12}\,\sin \theta_{13} \nonumber\\&&\cos(2\alpha+\delta+\phi_1+\phi_2)-\sin^2 \theta_{12} \sin \theta_{13} \cos(2\beta\nonumber\\&&+\delta+\phi_1+\phi_2)+\frac{b}{2} \sin 2 \theta_{12} \cos(2\alpha+\phi_1+\phi+3)\nonumber\\&&-\frac{1}{2}\sin 2\theta_{12}\cos (2\beta+\phi_1+\phi_3),
\end{eqnarray}
\begin{eqnarray}
B_2&=&\frac{1}{2} \sin 2\theta_{12}\big[-b \cos (2\alpha+\phi_1+\phi_2)+ \cos(2\beta+\phi_1+\nonumber\\&&\phi_2)\big]+a\sin \theta_{13} \cos (\delta-\phi_1-\phi_3) -b\cos^2 \theta_{12}\sin\theta_{13}\nonumber\\&&\cos (2\alpha+\delta+\phi_1+\phi_3)-\sin^2 \theta_{12} \sin\theta_{13}\cos (2\beta+\delta\nonumber\\&&+\phi_1+\phi_3),
\end{eqnarray}
\begin{eqnarray}
A_5 &=& \frac{1}{2} \sin 2\theta_{12}\big[-b\cos(2\alpha+\phi_1+\phi_3)+\sin\theta_{23}\cos(2\beta\nonumber\\&&+\phi_1+\phi_3)+b\cos\theta_{23}\sin(2\alpha+\phi_1+\phi_2)-\cos\theta_{23}\nonumber\\&&\sin(2\beta+\phi_1+\phi_2)\big],
\end{eqnarray}
\begin{eqnarray}
B_5&=& -a \cos\theta_{23}\cos(\delta-\phi_1-\phi_3)+b\cos^2 \theta_{12}\cos\theta_{23}\nonumber\\&&\cos(2\alpha+\delta+\phi_1+\phi_3)+\cos\theta_{23}\sin^2\theta_{12}\cos(2\beta\nonumber\\&&+\delta+\phi_1+\phi_3)+a \sin\theta_{23}\sin(\delta-\phi_1-\phi_2)+b\nonumber\\&&\cos^2 \theta_{12}\sin\theta_{23}\sin(2\alpha+\delta+\phi_1+\phi_2)+\sin^2 \theta_{12}\nonumber\\&&\sin\theta_{23}\sin(2\beta+\delta+\phi_1+\phi_2),
\end{eqnarray}
\begin{eqnarray}
A_6&=& a \sin\theta_{13}\sin\theta_{23} \cos(\delta-\phi_1-\phi_2)-b\cos^2\theta_{12} \sin\theta_{13}\nonumber\\&&\sin\theta_{23}\cos(2\alpha+\delta+\phi_1+\phi_2)-\sin^2\theta_{12}\sin\theta_{13}\nonumber\\&&\sin\theta_{23}\cos(2\beta+\delta+\phi_1+\phi_2)-\frac{1}{2}\sin2\theta_{12}\big[-b\nonumber\\&&sin\theta_{23}\sin(2\alpha+\phi_1+\phi_3)+\sin\theta_{23}\sin(2\beta+\phi_1\nonumber\\&&+\phi_3)\big],
\end{eqnarray}
\begin{eqnarray}
B_6&=& \frac{1}{2} \sin 2\theta_{12}\big[- b\cos(2\alpha+\phi_1+\phi_2)+\cos(2\beta+\phi_1\nonumber\\&&+\phi_2)\big]+a \sin\theta_{13}\sin(\delta-\phi_1-\phi_3)+b\cos^2\theta_{12} \nonumber\\&&\sin\theta_{13} \sin(2\alpha+\delta+\phi_1+\phi_3)+\sin^2 \theta_{12}\sin\theta_{13}\nonumber\\&&\sin(2\beta+\delta+\phi_1+\phi_3),
\end{eqnarray}
\begin{eqnarray}
B_3&=&-(\cos^2{\theta_{23}} \cos{(2(\alpha+\phi_2))} \sin^2{\theta_{12}}-\cos^2{\theta_{12}} \cos^2{\theta_{23}}\nonumber\\&& \cos{(2(\alpha+\delta+\phi_3))} \sin^2{\theta_{13}}+2\cos{\theta_{12}}\cos{\theta_{23}}\nonumber\\&&\cos{(2\alpha+\delta+2\phi_2)}\sin{\theta_{12}}\sin{\theta_{13}}\sin{\theta_{23}}+2\cos{\theta_{12}}\nonumber\\&&\cos{\theta_{23}}\cos{(2\alpha+\delta+2\phi_3)}\sin{\theta_{12}}\sin{\theta_{13}}\sin{\theta_{23}}-\nonumber\\&&\cos{(2(\alpha+\phi_3))} \sin^2{\theta_{12}} \sin^2{\theta_{23}} +\cos^2{\theta_{12}}\nonumber\\&&\cos{(2(\alpha+\delta+\phi_2))}\sin^2{\theta_{13}}\sin^2{\theta_{23}})(- \cos^2\theta_{13}\nonumber\\&& \sin^2\theta_{23}\sin2\phi_2 + \cos^2\theta_{13}\cos^2\theta_{23}\sin2\phi_3)+\nonumber\\&&(-\cos^2\theta_{13}\cos^2\theta_{23}\cos(2\phi_3) + \cos^2\theta_{13}\cos(2\phi_2)\nonumber\\&&\sin^2\theta_{23})(-\cos^2\theta_{23}\sin^2\theta_{12}\sin 2(\alpha+\phi_2)-\cos^2\theta_{12}\nonumber\\&&\sin^2\theta_{13}\sin^2\theta_{23}\sin 2(\alpha+\delta+\phi_2)-2\cos\theta_{12}\cos\theta_{23}\nonumber\\&&\sin\theta_{12}\sin\theta_{13}\sin\theta_{23}\sin(2\alpha+\delta+2\phi_2)+\sin^2\theta_{12}\nonumber\\&&\sin^2\theta_{23}\sin 2(\alpha+\phi_3)+\cos^2\theta_{12}\cos^2\theta_{23}\sin^2\theta_{13}\nonumber\\&&\sin 2(\alpha+\delta+\phi_3)-2\cos\theta_{12}\cos\theta_{23}\sin\theta_{12}\sin\theta_{13}\nonumber\\&&\sin\theta_{23}\sin(2\alpha+\delta+2\phi_3)),
\end{eqnarray}
\begin{eqnarray}
A_3&=&(-\cos^2\theta_{12}\cos^2 \theta_{23}\cos 2(\beta+\phi_2)+\cos^2 \theta_{23}\nonumber\\&& \cos 2(\beta+\delta+\phi_3)\sin^2\theta_{12}\sin^2\theta_{13}+2\cos\theta_{12}\nonumber\\&&\cos\theta_{23} \cos(2\beta+\delta+2\phi_2) \sin\theta_{12} \sin\theta_{13} \sin\theta_{23} \nonumber\\&& + 2 \cos\theta_{12} \cos\theta_{23} \cos(2\beta+\delta+2\phi_3) \sin\theta_{12}\nonumber\\&& \sin\theta_{13} \sin\theta_{23}+ \cos^2\theta_{12} \cos(2(\beta+\phi_3)) \sin^2\theta_{23}\nonumber\\&&- \cos(2(\beta+\delta+\phi_2)) \sin^2\theta_{12} \sin^2\theta_{13} \sin^2\theta_{23})\nonumber\\&&(-\cos^2{\theta_{23}}\sin^2{\theta_{12}}\sin{2(\alpha+\phi_2)}-\cos^2{\theta_{12}}\nonumber\\&&\sin^2{\theta_{13}}\sin^2{\theta_{23}}\sin{2(\alpha+\delta+\phi_2)}-2\cos{\theta_{12}}\nonumber\\&&\cos{\theta_{23}}\sin{\theta_{12}}\sin{\theta_{13}}\sin{\theta_{23}}\sin({2\alpha+\delta+2\phi_2})\nonumber\\&&+\sin^2{\theta_{12}}\sin^2{\theta_{23}}\sin{2(\alpha+\phi_3)}+\cos^2{\theta_{12}}\nonumber\\&&\cos^2{\theta_{23}}\sin^2{\theta_{13}}\sin{2(\alpha+\delta+\phi_3)}-2\cos{\theta_{12}}\nonumber\\&&\cos{\theta_{23}}\sin{\theta_{12}}\sin{\theta_{13}}\sin{\theta_{23}}\sin({2\alpha+\delta+2\phi_3}))\nonumber\\&&(\cos^2\theta_{23} \cos(2(\alpha+\phi_2))\sin^2\theta_{12} - \cos^2\theta_{12} \cos^2\theta_{23}\nonumber\\&& \cos(2(\alpha+\delta+\phi_3))\sin^2\theta_{13}+ 2\cos\theta_{12}\cos\theta_{23}\nonumber\\&&\cos(2\alpha+\delta+2\phi_2)\sin\theta_{12}\sin\theta_{13}\sin\theta_{23}+ 2\cos\theta_{12}\nonumber\\&&\cos\theta_{23}\cos(2\alpha+\delta+2\phi_3)\sin\theta_{12}\sin\theta_{13}\sin\theta_{23}- \nonumber\\&&\cos(2(\alpha+\phi_3))\sin^2\theta_{12}\sin^2\theta_{23}+ \cos^2\theta_{12}\cos(2(\alpha\nonumber\\&&+\delta+\phi_2))\sin^2\theta_{13}\sin^2\theta_{23})( -\cos^2\theta_{12}\cos^2\theta_{23}\nonumber\\&&\sin(2(\beta+\phi_2))
 - \sin^2\theta_{12}\sin^2\theta_{13}\sin^2\theta_{23}\sin(2(\beta\nonumber\\&&+\delta+\phi_2)) + 2\cos\theta_{12}\cos\theta_{23}\sin\theta_{12}\sin\theta_{13}\sin\theta_{23}\nonumber\\&&\sin(2\beta+\delta+2\phi_2) + \cos^2\theta_{12}\sin^2\theta_{23}\sin(2(\beta+\phi_3))\nonumber\\&& + \cos^2\theta_{23}\sin^2\theta_{12}\sin^2\theta_{13}\sin(2(\beta+\delta+\phi_3)) + \nonumber\\&&2\cos\theta_{12}\cos\theta_{23}\sin\theta_{12}\sin\theta_{13}\sin\theta_{23}\sin(2\beta+\delta\nonumber\\&&+2\phi_3)),
\end{eqnarray}

\begin{eqnarray}
A_4&=& -\cos^2\theta_{12} \cos^2\theta_{23} \cos2(\beta + \phi_2) \sin^2\theta_{23} \sin2\phi_2 \nonumber\\&&+\cos^2\theta_{23} \cos2(\beta + \delta + \phi_3) \sin^2\theta_{12}\sin^2\theta_{13}\nonumber\\&&  \sin^2\theta_{23} \sin2\phi_2 + 2 \cos\theta_{12} \cos\theta_{23} \cos(2\beta + \delta \nonumber\\&& + 2\phi_2) \sin\theta_{12}\sin\theta_{13} \sin^3\theta_{23} \sin2\phi_2 +2 \cos\theta_{12}\nonumber\\&& \cos\theta_{23} \cos(2\beta + \delta + 2\phi_3) \sin\theta_{12} \sin\theta_{13} \sin^3\theta_{23} \nonumber
\end{eqnarray}
\begin{eqnarray} 
&=&\sin2\phi_2 +\cos^2\theta_{12} \cos2(\beta + \phi_3) \sin^4\theta_{23} \sin2\phi_2 \nonumber\\&& - 
\cos2(\beta + \delta + \phi_2) \sin^2\theta_{12} \sin^2\theta_{13} \sin^4\theta_{23} \sin2\phi_2 \nonumber\\&&-\cos^2\theta_{12} \cos^2\theta_{23} \cos2\phi_3 \sin 2(\beta + \phi_2)+\nonumber\\&&\cos^2{\theta_{12}}\cos^2{\theta_{23}}\cos{2\phi_2}\sin^2{\theta_{23}}\sin{2(\beta+\phi_2)}\nonumber\\&&- \cos{\theta_{23}}^2\cos{2\phi_3}\sin^2{\theta_{12}}\sin^2{\theta_{13}}\sin^2{\theta_{23}}\nonumber\\&&\sin{2(\beta+\delta+\phi_2)}+ \cos{2\phi_2}\sin^2{\theta_{12}}\sin^2{\theta_{13}}\nonumber\\&&\sin^2{\theta_{23}}\sin{2(\beta+\delta+\phi_2)} + 2\cos{\theta_{12}}\cos^3{\theta_{23}}\nonumber\\&&\cos{2\phi_3}\sin{\theta_{12}}\sin{\theta_{13}}\sin{\theta_{23}}\sin{2\beta+\delta+2\phi_2}\nonumber\\&&- 2\cos{\theta_{12}}\cos{\theta_{23}}\cos{2\phi_2}\sin{\theta_{12}}\sin{\theta_{13}}\sin^3{\theta_{23}}\nonumber\\&&\sin{2\beta+\delta+2\phi_2} + \cos^2{\theta_{12}}\cos^4{\theta_{23}}\cos{2(\beta+\phi_2)}\nonumber\\&&\sin{2\phi_3}- \cos^4{\theta_{23}}\cos{2(\beta+\delta+\phi_3)}\sin^2{\theta_{12}}\nonumber\\&&\sin^2{\theta_{13}}\sin{2\phi_3}- 2\cos{\theta_{12}}\cos^3{\theta_{23}}\cos(2\beta+\nonumber\\&&\delta+2\phi_2)\sin{\theta_{12}}\sin{\theta_{13}}\sin{\theta_{23}}\sin{2\phi_3}-2 \cos\theta_{12} \nonumber\\&&
\cos^3\theta_{23} \cos(2\beta + \delta + 2\phi_3) \sin\theta_{12} \sin\theta_{13} \sin\theta_{23}\nonumber\\&& \sin(2\phi_3)- \cos^2\theta_{12} \cos^2\theta_{23} \cos(2(\beta+\phi_3)) \sin^2\theta_{23} \nonumber\\&&\sin(2\phi_3)+ \cos^2\theta_{23} \cos(2(\beta + \delta + \phi_2)) \sin^2\theta_{12} \nonumber\\&&\sin^2\theta_{13} \sin^2\theta_{23} \sin(2\phi_3)+ \cos^2\theta_{12} \cos^2\theta_{23} \cos(2\phi_3)\nonumber\\&& \sin^2\theta_{23} \sin(2(\beta+\phi_3))- \cos^2\theta_{12} \cos(2\phi_2) \sin^4\theta_{23}\nonumber\\&& \sin(2(\beta+\phi_3))+ \cos^4\theta_{23} \cos(2\phi_3) \sin^2\theta_{12} \sin^2\theta_{13} \nonumber\\&&\sin(2(\beta+\delta+\phi_3))- \cos^2\theta_{23} \cos(2\phi_2) \sin^2\theta_{12} \nonumber\\&&\sin^2\theta_{13} \sin^2\theta_{23} \sin(2(\beta+\delta+\phi_3)) + 2 \cos\theta_{12}\nonumber\\&& \cos^3\theta_{23} \cos(2\phi_3) \sin\theta_{12} \sin\theta_{13} \sin\theta_{23} \sin(2\beta + \nonumber\\&&\delta + 2\phi_3)-2 \cos\theta_{12} \cos\theta_{23} \cos(2\phi_2) \sin\theta_{12}\nonumber\\&& \sin\theta_{13} \sin^3\theta_{23} \sin(2\beta + \delta + 2\phi_3),
\end{eqnarray}

\begin{eqnarray}
B_4&=& \cos^2\theta_{23}\cos2\left(\alpha+\phi_2\right)\sin^2\theta_{12}\sin^2\theta_{23}\nonumber\\&&\sin2\phi_2 -\cos^2\theta_{12}\cos^2\theta_{23}\cos2\left(\alpha+\delta+\phi_3\right)\nonumber\\&&\sin^2\theta_{13}\sin^2\theta_{23}\sin2\phi_2+2\cos\theta_{12}\cos\theta_{23}\nonumber\\&&\cos2\alpha+\delta+2\phi_2\sin\theta_{12}\sin\theta_{13}\sin^3\theta_{23}\sin2\phi_2\nonumber\\&& +2\cos\theta_{12}\cos\theta_{23}\cos2\alpha+\delta+2\phi_3\sin\theta_{12}\sin\theta_{13}\nonumber\\&&\sin^3\theta_{23}\sin2\phi_2-\cos2\left(\alpha+\phi_3\right)\sin^2\theta_{12}\sin^4\theta_{23}\nonumber\\&&\sin2\phi_2+\cos^2\theta_{12}\cos2\left(\alpha+\delta+\phi_2\right)\sin^2\theta_{13}\nonumber\\&&\sin^4\theta_{23}\sin2\phi_2+\cos^4\theta_{23}\cos2\phi_3\sin^2\theta_{12}\nonumber\\&&\sin^2\left(\alpha+\phi_2\right)-\cos^2\theta_{23}\cos2\phi_2\sin^2\theta_{12}\sin^2\theta_{23}\nonumber\\&&\sin2\left(\alpha+\phi_2\right)+\cos^2\theta_{12}\cos^2\theta_{23}\cos2\phi_3\sin^2\theta_{13}\nonumber\\&&\sin^2\theta_{23}\sin2\left(\alpha+\delta+\phi_2\right)-\cos^2\theta_{12}\cos(2\phi_2)\nonumber\\&&\sin^2\theta_{13}\sin^4\theta_{23}\sin(2(\alpha+\delta+\phi_2))+2\cos\theta_{12}\nonumber\\&&\cos^3\theta_{23}\cos(2\phi_3)\sin\theta_{12}\sin\theta_{13}\sin\theta_{23}\sin(2\alpha+\nonumber\\&&\delta+2\phi_2)-2\cos\theta_{12}\cos\theta_{23}\cos(2\phi_2)\sin\theta_{12}\sin\theta_{13}\nonumber\\&&\sin^3\theta_{23}\sin(2\alpha+\delta+2\phi_2)-\cos^4\theta_{23}\cos2(\alpha+\phi_2)\nonumber\\&&\sin^2\theta_{12}\sin(2\phi_3)+\cos^2\theta_{12}\cos^4\theta_{23}\cos(2(\alpha+\delta+\nonumber\\&&\phi_3))\sin^2\theta_{13}\sin(2\phi_3)-2\cos\theta_{12}\cos^3\theta_{23}\cos(2\alpha+\nonumber\\&&\delta+2\phi_2)\sin\theta_{12}\sin\theta_{13}\sin\theta_{23}\sin(2\phi_3)-2\cos\theta_{12}\nonumber\\&&\cos^3\theta_{23}\cos(2\alpha+\delta+2\phi_3)\sin\theta_{12}\sin\theta_{13}\sin\theta_{23}\nonumber\\&&\sin(2\phi_3)+\cos^2\theta_{23}\cos(2(\alpha+\phi_3))\sin^2\theta_{12}\sin^2\theta_{23}\nonumber
\end{eqnarray}
\begin{eqnarray}
&=&\sin(2\phi_3)-\cos^2{\theta_{12}}\cos^2{\theta_{23}}\cos2(\alpha+\delta+\phi_2)\nonumber\\&&\sin^2{\theta_{13}}\sin^2{\theta_{23}}\sin2\phi_3-\cos^2{\theta_{23}}\cos2\phi_3\sin^2{\theta_{12}}\nonumber\\&&\sin^2{\theta_{23}}\sin2(\alpha+\phi_3)+\cos2\phi_2 \sin^2{\theta_{12}}\sin^4{\theta_{23}}\nonumber\\&&\sin2(\alpha+\phi_3)-\cos^2{\theta_{12}}\cos^4{\theta_{23}}\cos2\phi_3\sin^2{\theta_{13}}\nonumber\\&&\sin2(\alpha+\delta+\phi_3)+\cos^2{\theta_{12}}\cos^2{\theta_{23}}\cos2\phi_2\nonumber\\&&\sin^2{\theta_{13}}\sin^2{\theta_{23}}\sin2(\alpha+\delta+\phi_3)+2\cos{\theta_{12}}\nonumber\\&&\cos^3{\theta_{23}}\cos2\phi_3\sin{\theta_{12}}\sin{\theta_{13}}\sin{\theta_{23}}\sin(2\alpha+\delta\nonumber\\&&+2\phi_3)-2\cos{\theta_{12}}\cos{\theta_{23}}\cos2\phi_2\sin{\theta_{12}}\sin{\theta_{13}}\nonumber\\&&\sin^3{\theta_{23}}\sin2\alpha+\delta+2\phi_3.
\end{eqnarray}

\end{document}